\documentclass[aps,prl,reprint,showpacs,superscriptaddress]{revtex4-1}
\usepackage{graphicx}
\usepackage{dcolumn}
\usepackage{bm}
\begin{document}


\title{Generation of high-energy monoenergetic heavy ion beams by radiation pressure acceleration of ultra-intense laser pulses}


\author{D. Wu}
\affiliation{Center for Energy Research, University of California, San Diego, California 92093, USA.}
\affiliation{Key Laboratory of HEDP of the Ministry of Education, CAPT, and State Key Laboratory of Nuclear Physics and Technology, Peking University, Beijing, 100871, China.}
\author{B. Qiao}
\email{bqiao@ucsd.edu}
\affiliation{Center for Energy Research, University of California, San Diego, California 92093, USA.}
\affiliation{Key Laboratory of HEDP of the Ministry of Education, CAPT, and State Key Laboratory of Nuclear Physics and Technology, Peking University, Beijing, 100871, China.}
\author{X. T. He}
\affiliation{Key Laboratory of HEDP of the Ministry of Education, CAPT, and State Key Laboratory of Nuclear Physics and Technology, Peking University, Beijing, 100871, China.}
\author{C. McGuffey}
\affiliation{Center for Energy Research, University of California, San Diego, California 92093, USA.}
\author{F. N. Beg}
\affiliation{Center for Energy Research, University of California, San Diego, California 92093, USA.}


\date{\today}

\begin{abstract}

A novel radiation pressure acceleration (RPA) regime of heavy ion beams from laser-irradiated ultrathin foils is proposed by self-consistently taking into account the ionization dynamics. In this regime, the laser intensity is required to match with the large ionization energy gap when the successive ionization of high-Z atoms passing the noble gas configurations [such as removing an electron from the helium-like charge state $(\text{Z}-2)^+$ to $(\text{Z}-1)^+$].  While the target ions in the laser wing region are ionized to low charge states and undergo rapid dispersions due to instabilities, a self-organized, stable RPA of highly-charged heavy ion beam near the laser axis is achieved. It is also found that a large supplement of electrons produced from ionization helps preserving stable acceleration. Two-dimensional particle-in-cell simulations show that a monoenergetic $\text{Al}^{13+}$ beam with peak energy $1\ \text{GeV}$ and energy spread of $5\%$ is obtained by lasers at intensity $7\times10^{20}\ \text{W}/\text{cm}^2$. 
   
\end{abstract}

\pacs{52.38.Kd, 41.75.Jv, 52.35.Mw, 52.59.-f}

\maketitle

\textit{Introduction --} Laser driven ion acceleration has become a highly active field of research over the past decade years\cite{Rev.Mod.Phys.85.751,Daido12}. The wide potential applications \cite{Borghesi06, Malka08} include tumor therapy, ultrafast radiography, isotope production and fast ignition inertial confinement fusion. Most of the applications require a high-energy ion beam with large particle number, monoenergetic spectrum and small angular divergence. Among a number of acceleration schemes identified to date, including target normal sheath acceleration (TNSA) \cite{Wilks01,Mackinnon02}, collisionless shock acceleration \cite{Haberberger12} and others, the newly-emerged radiation pressure acceleration (RPA) \cite{Macchi05,Robinson08,Qiao09,Macchi09, Qiao10} using circularly polarized lasers has been regarded as the most promising route to obtaining such high-quality ion beams in a much more efficient manner. However, all the current studies are focused on protons and light ions. How to use RPA to produce high-energy monoenergetic heavy ion beams is still unknown. Meanwhile, comparing to proton and light ions, heavy ion beams have a much broader range of applications related to medicine, materials \cite{Habs11}, nuclear fission and fusion \cite{Beg02, McKenna04}, Quantum Electrodynamics \cite{Stöhlker03} and High Energy Density Plasmas \cite{Logan08}.

In the idealistic RPA scheme all ion species are co-accelerated and the heavy ion acceleration is not impeded by the light ion contaminants (hydrogen, carbon and oxygen) \cite{Robinson08,Macchi09}, which are different from the widely studied TNSA mechanism. So in principle heavy ions can be efficiently accelerated to much higher energy in RPA due to their large mass. However, both simulations and experiments \cite{Qiao11,Kar12} show that in a more realistic multi-dimensional case, the heavy ion species at later time undergo rapid Coulomb explosion leading to very broad energy spectrum due to the deficiency of co-moving electrons and the transverse instabilities. Recently, two-dimensional (2D) particle-in-cell (PIC) simulation \cite{Korzhimanov12} indicates that a quasi-monoenergetic high-energy $\text{Fe}^{24+}$ beam can be generated by using compound targets mixed with heavier Au substrate. However, it relies on complicated target fabrication and the laser-ion conversion efficiency is certainly extremely low.  

In this Letter, we propose a new RPA regime for generation of high-energy monoenergetic heavy ion beams from ultrathin foil targets by intense laser pulses, in which the ionization dynamics of high-Z atoms are self-consistently taken into account. Basically, with an appropriate matching condition between laser intensity, foil thickness and the unique large gap feature in the ionization energies when the successive ionization of high-Z atoms passing the noble gas configuration, such as removing an electron from the helium-like charge state $(\text{Z}-2)^+$ to $(\text{Z}-1)^+$, a self-organized, stable RPA of highly-charged heavy ion beam close to the laser central axis can be obtained. The stabilization is also related to a large supplement of electrons produced from ionization, which keeps the accelerating plasma nontransparent to laser and suppresses the beam transverse instabilities. The condition for this regime has been analytically established and verified by our 2D PIC simulations. They show that a highly-charged ($13+$) monoenergetic aluminum beam with total charge above 
$2\ \text{nC}$, peak energy $1\ \text{GeV}$, and energy spread only $5\%$ is produced by intense circularly-polarized laser at intensity $7\times10^{20}\ \text{W}/\text{cm}^2$ through the regime. Similarly, monoenergetic $\text{Fe}^{26+}$ beam with peak energy $16.8\ \text{GeV}$ can also be obtained 
if the laser intensity is increased to $4\times10^{22}\ \text{W}/\text{cm}^2$.

\textit{Heavy ion ionization gap feature --} When an intense laser pulse irradiates a solid high-Z foil target, a crucial, but unexplored yet effect for RPA of heavy ions that needs to be considered is the successive ionization process, which includes both optical field \cite{Kemp04} and impact ionizations \cite{Sentoku08}.  In the Light-sail RPA scheme with ultrathin foils, both electrons and ions are co-accelerated to high energy (i.e. high temperature) with almost zero relative velocity, the optical field ionization dominates. For high-Z atoms in a electric field, the ionization rate from $(\zeta-1)^+$ to $\zeta^+$ ($\zeta\leq\text{Z}$) charge states can be calculated from the direct current Ammosov-Delone-Krainov (ADK) formula \cite{Ammosov}:
\begin{eqnarray}
\label{Eq:ionizationrate}
&& W_{l,m}=\omega_aC^2_{n^{\ast}l^{\ast}}
\frac{(2l+1)(l+\vert{m}\vert)!}{2^{\vert{m}\vert}(\vert{m}\vert)!(l-\vert{m}\vert)!}\frac{U_{\text{i}}}{2U_{\text{H}}} \nonumber \\
&& \times[\frac{2E_{\text{H}}}{E}(\frac{U_{\text{i}}}{U_{\text{H}}})^{3/2}]^{2n^{\ast}-\vert{m}\vert-1}
 \times\exp[-\frac{2}{3}\frac{E_{\text{H}}}{E}(\frac{U_{\text{i}}}{U_{\text{H}}})^{3/2}],
\end{eqnarray}
where $\omega_a=\alpha^3c/r_e=4.13\times10^{-16}\ \text{s}^{-1}$ is the atomic unit frequency, $U_{\text{i}}$ is the ion ionization potential, and $U_{\text{H}}=1312\ \text{kJ}/\text{mol}$ is the ionization potential of Hydrogen (H) at the fundamental state, $E_{\text{H}}=m_e^2e^5h^{-4}=5.14\times10^{11}\ \text{V}/\text{m}$ is the corresponding electric field, $m_e$ is the electron mass, $c$ is the speed of light, $\alpha=1/137$ is the fine structure constant, $r_e=m_e/c^2$ is the classical electron radius, $l$, $m$ are the electrons' orbital quantum number and its projection, respectively, $n^{\ast}=\zeta\sqrt{U_{\text{H}}/U_{\text{i}}}$ is the effective principal quantum number, $l^{\ast}=n_0^{\ast}-1$ is the effective value of the orbital number, $n_0^{\ast}$ is the effective principal quantum number of the ground state, and $\zeta$ ($\zeta\leq\text{Z}$) is the ion charge number after ionization. The coefficients $C_{n^{\ast}l^{\ast}}=2^{2n^{\ast}}/[n^{\ast}\Gamma(n^{\ast}+l^{\ast}+1)\Gamma(n^{\ast}-l^{\ast})]$ in Eq. (\ref{Eq:ionizationrate}) can be calculated using semi-classical approximation as $C_{n^{\ast}l^{\ast}}\simeq1/2\pi n^{\ast}\times[4e^2/(n^{\ast2}-l^{\ast2}]^{n^{\ast}}\times[(n^{\ast}-l^{\ast})(n^{\ast}+l^{\ast})]^{l^{\ast}+1/2}$. In the limit $l^{\ast}<<n^{\ast}$, considering the fundamental state $l=m=0$, we obtain the ionization rate as a function of the local electric field for the charge state at $\zeta^+$as,
\begin{equation}
\label{Eq:rate}
W=\frac{\omega_a}{8\pi}\frac{E}{E_{\text{H}}\zeta}\times(\frac{4\pi E_{\text{H}} \zeta^3}{E n^{\ast4}})^{2n^{\ast}}\times\exp[-\frac{2}{3}\frac{E_{\text{H}}}{E}(\frac{\zeta}{n^{\ast}})^3].
\end{equation}

\begin{figure}
\includegraphics[width=8.50cm]{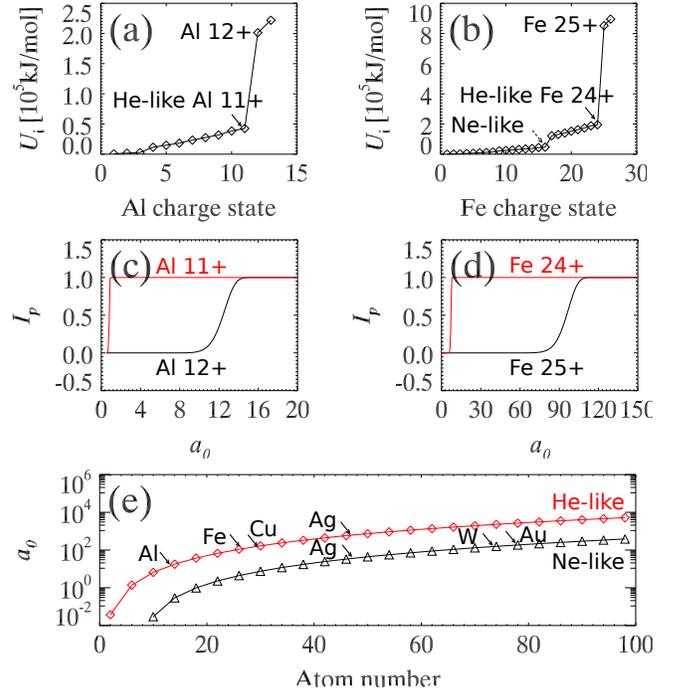}
\caption{\label{fig:ionization} (color online) (a) and (b) the ionization energy potentials vs ion charge state $\zeta$ for respectively Al and Fe.
(c) and (d) the ionization probabilities of $\text{Al}^{11+}$, $\text{Al}^{12+}$ and $\text{Fe}^{24+}$, $\text{Fe}^{25+}$ vs normalized laser electric field (amplitude) $a_0=eE/m_ec\omega_0$. (e) The required laser amplitude $a_0$ to achieve stable heavy ion RPA regime vs the atomic number Z of different target materials, which is calculated from Eqs. (\ref{Eq:rate})-(\ref{Eq:He}) by matching with the He-like ionization potential gaps. Those for matching with the Ne-like gaps are also plotted.}
\end{figure}

The ionization probability within a finite time $\delta t$ is calculated as 
\begin{equation}
\label{Eq:ionizationprobablity}
I_p=1-\exp(-W \delta t).
\end{equation}
From Eqs. (\ref{Eq:rate}) and (\ref{Eq:ionizationprobablity}), we see that the ionization dynamics heavily depends on the ionization potentials of different charge states for different materials. Figures \ref{fig:ionization}(a) and \ref{fig:ionization}(b) plot the ionization energy potentials in different charge states for respectively Aluminum (Al) and iron (Fe) elements based on the NIST database \cite{NIST}. It can be clearly seen that a large gap in the ionization energies exists when the successive ionization of high-Z atoms passing the noble gas configuration, such as when the ionization passing the helium (He) -like configuration for Al from $\text{Al}^{11+}$ to $\text{Al}^{12+}$ and for Fe from $\text{Fe}^{24+}$ to $\text{Fe}^{25+}$ [generally the charge state increases from $\zeta^+=(Z-2)^+$ to $(Z-1)^+$]. In fact, such large ionization potential gap features are general for high-Z atoms because atoms/ions with nobel-like configurations are always rather stable, which includes not only He-like but also Neon (Ne) -like. The ionization potentials for He and Ne-like configuration of atoms/ions varying with the target material atomic number Z can be found from the NIST data base \cite{NIST} as well. And an empirical formula describing the dependence of the ionization potential of He-like ions on their atomic number Z is given as \cite{Natur.94.779},
\begin{equation} 
\label{Eq:He}
U_{\text{He}\text{-}\text{like}}=\frac{65}{64}(Z^2-\frac{4}{3}Z+\frac{4}{9})\times U_{\text{H}},
\end{equation}
where $U_{\text{H}}=1316$ kJ$/$mol is the ionization potential of H atom. 

\begin{figure*}
\includegraphics[width=16.0cm]{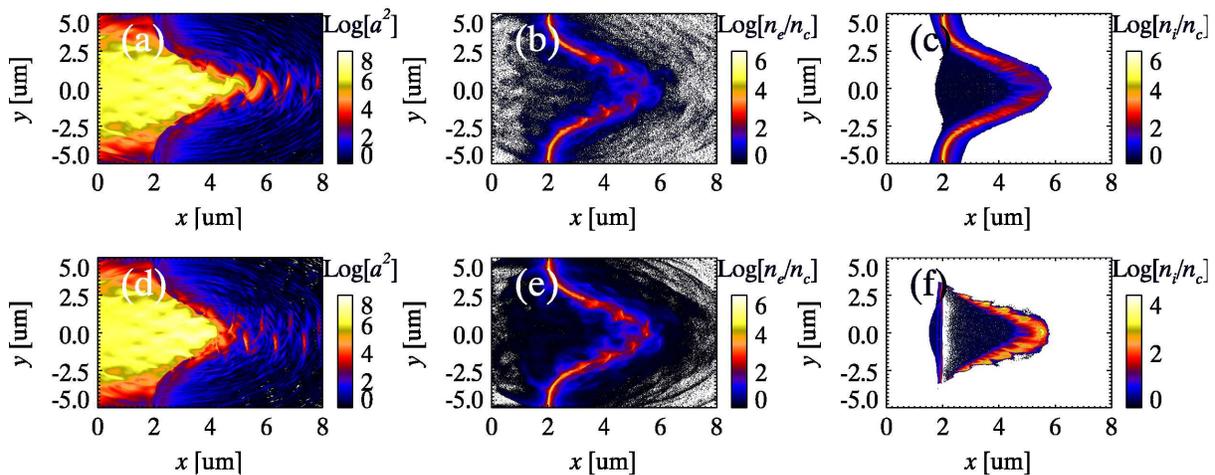}
\caption{\label{fig:Alacceleration} (color online) 2D PIC simulation results: laser intensity [(a) and (d)] at $t=25T_0$, electron [(b) and (e)] and ion [(c) and (f)] densities at $t=30T_0$ for ultrathin $20\ \text{nm}$ Al foils irradiated by CP lasers at intensity $7\times10^{20}\ \text{W}/\text{cm}^2$, where the ionization effect is not taken into account in (a)-(c) but self-consistently included in (d)-(f). The other parameters are shown in the text.}
\end{figure*}

\textit{Novel regime of stable heavy ion RPA --} Our stable heavy ion RPA regime is based on the above ionization gap feature of high-Z ultrathin foil targets during their interaction with intense laser pulses. According to Eqs. (\ref{Eq:rate})-(\ref{Eq:He}), the large gaps in the ionization potentials with high-Z atoms indicate that a sharp increase of the electric field is required to achieve the highly-charged state of heavy ions, such as $(\text{Z}-1)^+$ of one-electron system. If we assume the laser electric field dominates during the successive ionization, Figures \ref{fig:ionization}(c) and \ref{fig:ionization}(d) plot the calculated ionization probability $I_p$ of respectively $\text{Al}^{11+}$, $\text{Al}^{12+}$ and $\text{Fe}^{24+}$, $\text{Fe}^{25+}$ varying with the normalized laser electric field, i.e. the laser amplitude $a_0=eE/m_ec\omega_0$, where $e$ and $m_e$ are electron charge and mass, $\omega_0$ is the laser frequency. $\delta t$ is chosen to be equal to the laser pulse duration $\tau$ and $\tau=30T_0$ ($T_0=2\pi /\omega_0$) is assumed here. We see that  [Fig. \ref{fig:ionization}(c)]  for a large range of laser amplitude $1<a_0<12$, the dominant charge state of Al ions in the laser-driven acceleration is $\text{Al}^{11+}$, while to achieve the highly-charged state $\text{Al}^{12+}$ dominating, the laser field has to be sharply increases from $a_0=12$ to $a_0>16$. Similar feature can also be seen in Fe in Fig. \ref{fig:ionization}(d), where to achieve $\text{Fe}^{25+}$, $a_0>120$ is required. 

To achieve stable RPA of heavy ions, one possible way is to choose the peak laser amplitude $a_0$ to be closely matching to that required to overcome the above ionization gap and achieve highly-charged state $(\text{Z}-1)^+$ of heavy ion species. This will help to stabilize the acceleration in two ways. Firstly, since intense laser amplitude generally has transverse Gaussian distribution as $a=a_0*exp(-r^2/r_0^2)$ (where $r_0$ is the laser spot radius), only the ions near the central laser axis can be ionized to extremely high charge state $(\text{Z}-1)^+$ (or $\text{Z}^+$), which undergo strong acceleration by the larger electric force due to their higher charge to mass ratio. The ions in the laser wing region are ionized to lower charge states due to dropping of the laser field, undergoing weaker acceleration and rapid dispersion due to their lower charge to mass ratio and the transverse instabilities. Eventually, a self-organized, transverse-size-limited ion beam near the laser axis is formed undergoing efficient acceleration, where the one-dimensional (1D) stable RPA dynamics is maintained. Secondly and most importantly, the high-order ionization process occurring near the laser axis will produce a large number of supplementary electrons accompanying the heavy ion beam acceleration, which would helps to avoid Coulomb explosion and preserve stable RPA of the heavy ion beam in the center, as expected from Refs. \cite{Qiao10, Qiao11pop}. Therefore, a high-energy monoenergetic highly-charged heavy ion beams can be generated in this new RPA regime. The required laser amplitude $a_0$ for this stable heavy ion RPA regime can be calculated from the coupled equations (\ref{Eq:rate})-(\ref{Eq:He}), which is shown in Fig. \ref{fig:ionization}(e). The other condition for the regime is that the foil thickness should satisfy the optimal condition of RPA \cite{Macchi09}, that is, the laser radiation pressure balances the maximum charge separation electric field force $l_0/\lambda\simeq a_0n_c/\pi Z_{eff}n$, where $Z_{eff}$ is the effective charge states of high-Z targets during acceleration.
 
\textit{Verification by PIC simulations --} In order to verify the new regime discussed above, 2D PIC simulations are carried out, where the field ionization module is self-consistently integrated in the code. Multi-level ionization and energy conservation during the ionization process has also been taken into account.  An ultrathin Al foil target with solid density $\rho_{\text{Al}}=2.7\ \text{g}/\text{cm}^3$ (atom density $6\times10^{22}\ \text{cm}^{-3}$) is chosen. According to the conditions discussed above [Fig. \ref{fig:ionization}(e)], a circularly polarized (CP) laser pulse with amplitude $a_0=16$ (intensity $7\times10^{20}\ \text{W}/\text{cm}^2$) and wavelength $\lambda=1\ \mu\text{m}$ is taken to be normally incident on the Al foil. The laser pulse has a transverse Gaussian profile with spot radius $r_0=3.0\ \mu\text{m}$ and a trapezoidal temporal profile of duration $\tau=30T_0$, consisting of a plateau of $26T_0$ and rising and falling times of $2T_0$ each. The initial foil thickness is chosen to be $20\ \text{nm}$ satisfying the above optimal RPA condition and the initial charge state is taken as $\text{Al}^{3+}$ as the average charge state under 
room temperature $T_e=T_{\text{Al}}=10\ \text{eV}$. In the simulations, 8000 cells along the x axis and 1200 cells transversely along the y axis constitute a $8\times12\ \mu\text{m}$ simulation box. Each foil cell is filled up with 900 electrons and 900 $\text{Al}^{3+}$ ions. The Al foil is located at $x=2.0\ \mu\text{m}$ and the laser propagates from the left boundary of the simulation box. To more clearly identify the role of the ionization on the RPA dynamics of heavy ion beams, the simulations without the ionization effect included are also carried out for comparison, where the charge state is fixed to $\text{Al}^{11+}$.

\begin{figure}\label{3}
\includegraphics[width=8.50cm]{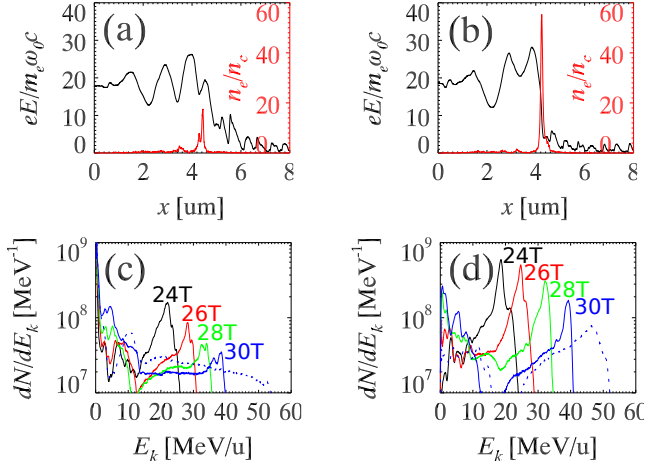}
\caption{\label{fig:Alspectra} (color online) The corresponding longitudinal profiles of laser electric field $E$ and electron density $n_e/n_c$ at $y=0$ and $t=25T_0$ in the simulation of Fig. \ref{fig:Alacceleration} for the cases of respectively without (a) and with (b) ionization effects. (c) and (d) are respectively the energy spectra of Al ions at time $t=24$, $26$, $28$ and $30T_0$. Note that in (d) at $t=24$ and $26T_0$, the spectra are calculated for the combination of both $\text{Al}^{12+}$ and $\text{Al}^{13+}$. The dashed lines in (c) and (d) show the results when laser pulse temporal profile changes to be $\sin^2$ function.}
\end{figure}

Figure \ref{fig:Alacceleration} shows the laser intensity distributions at time $t=25T_0$ and the electron and Al ion density maps at $t=30T_0$ for respectively without [(a)-(c)] and with [(d)-(f)] the ionization effect taken into account.  From Fig. \ref{fig:Alacceleration}(a), we can see that without ionization the intense laser pulse easily penetrates through the accelerating plasma slab due to transverse instabilities, leading to rapid heating and dispersion of the electrons in the latter [see Fig. \ref{fig:Alacceleration}(b)], which eventually results in decompression and termination of the accelerating $\text{Al}^{11+}$ ion beam. However, when the ionization is taken into account and the above condition for our heavy ion RPA regime are satisfied, as expected, a self-organized highly-charged $\text{Al}^{13+}$ ion beam near the laser axis is formed at later time, which undergoes stable RPA by the intense laser pulse, as shown in Fig. \ref{fig:Alacceleration}(f). Furthermore, comparing Figs. \ref{fig:Alspectra}(a) with \ref{fig:Alspectra}(b), we clearly see that the ionization produces a large number of additional electrons as a supplementary to the accelerating plasma slab, where the electron density is much higher than the case without ionization. This keeps the accelerating plasma slab being nontransparent to intense laser field and helps to stabilize the RPA of heavy $\text{Al}^{13+}$ ion beam, as we mentioned above and in Refs. \cite{Qiao10, Qiao11pop}. 

The energy spectra of heavy Al ions for both cases at time $t=24$, $26$, $28$ and $30T_0$ are shown in Figs. \ref{fig:Alspectra}(c) and \ref{fig:Alspectra}(d) respectively. Clearly without the ionization effect, the heavy $\text{Al}^{11+}$ ion beam initially has a well-peaked energy spectrum at $t=24T_0$ [the black line in \ref{fig:Alspectra}(c)], but at later time it undergoes rapid instabilities and Coulomb explosion, leading to much broadened energy spectrum without any obvious peak finally of the beam at $t=30T_0$ [the blue line in \ref{fig:Alspectra}(c)]. However, in our regime with the ionization taken into account [Fig. \ref{fig:Alspectra}(d)], the heavy Al ion beam experience a stable RPA, the pronounced peak in the energy spectrum is preserved till the laser is over at $t=30T_0$. On the other hand, because the ionization potential difference between $\text{Al}^{12+}$ and $\text{Al}^{13+}$ is very small, the
ion beam at $t=24$ and $26T_0$ [the black and red lines in \ref{fig:Alspectra}(d)] has both $\text{Al}^{12+}$ and $\text{Al}^{13+}$ charge states, which are further rapidly ionized to fully $\text{Al}^{13+}$ at later time [the green and blue lines in \ref{fig:Alspectra}(d)]. Eventually at $t=30T_0$ when the laser pulse is over, a quasi-monoenergetic $\text{Al}^{13+}$ ion beam with peak energy of $40\ \text{MeV}/\text{u}$ ($1.08\ \text{GeV}$) and energy spread of only $5\%$ is produced. The effective particle number in the beam is about $10^9$, which means the total charge about $2\ \text{nC}$.

\begin{figure}\label{4}
\includegraphics[width=8.50cm]{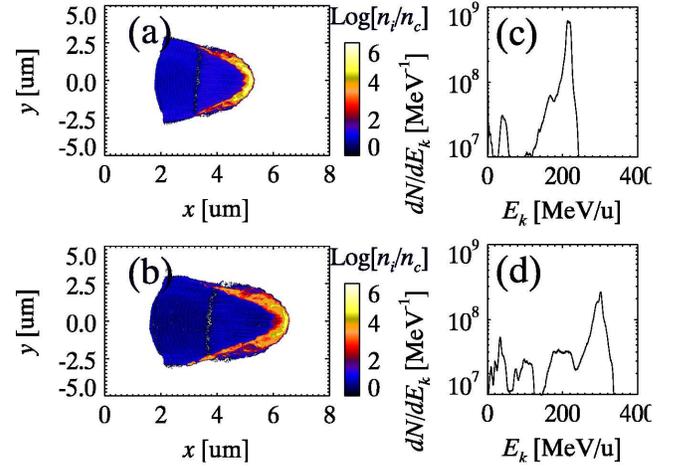}
\caption{\label{fig:Feacceleration} (color online) 2D PIC simulation results for ultrathin Fe foil:  ion density [(a) and (b)] and energy spectrum [(c) and (d)] for $\text{Fe}^{26+}$ ions at $t=12$ and $14T_0$ respectively, where the foil thickness is $30\ \text{nm}$, the laser amplitude is $a_0=130$ ($10^{22}\ \text{W}/\text{cm}^2$) and pulse length is $\tau=14T_0$. Other parameters are the same as those in Fig. \ref{fig:Alacceleration}.}
\end{figure}  

In order to identify the robustness of the above new heavy ion RPA regime, further simulations with Fe ultrathin foil targets are also carried out. To match the condition of the regime [Eqs. (\ref{Eq:rate})-(\ref{Eq:He})], the incident CP laser amplitude is chosen to be $a_0=130$ (intensity $I\simeq4\times10^{22}\ \text{W}/\text{cm}^2$) and the pulse duration is $\tau=14T_0$. The thickness of the solid Fe foil at density $7.9\ \text{g}/\text{cm}^3$ is taken as $30\ \text{nm}$ and the initial charge state is set to be $\zeta_0=3$ as well. All other parameters are the same as above. Figure \ref{fig:Feacceleration} shows density maps and energy spectra of $\text{Fe}^{26+}$ ions at $t=12$ and $14T_0$ respectively. It can be clearly seen that a quasi-monoenergetic $\text{Fe}^{26+}$ ion beam with peak energy $300\ \text{MeV}/\text{u}$ ($16.8\ \text{GeV}$) is produced. Note that such extremely high-energy heavy ion beams can be even used for creation of quark-gluon plasmas \cite{Abbott00}. For super heavy ion beams, such as W and Au, in order to obtain stable RPA of highly-charge $\text{W}^{73+}$ and $\text{Au}^{78+}$ [$(\text{Z}-1)^+$] ion beams, the corresponding laser intensities need to be as high as $10^{25}\ \text{W}/\text{cm}^2$ to match their He-like ionization potential gaps, which are unachievable so far. However, we can still carefully match the laser intensities with their Ne-like ionization potential gaps to achieve stable RPA of lower-charge state such as $\text{W}^{65+}$ and $\text{Au}^{70+}$ [$(\text{Z}-9)^+$], where the required laser intensity would drop to the order of $10^{22}\ \text{W}/\text{cm}^2$, reachable with current laser facilities. 

\textit{Discussion --} Note that in the above simulations, the reason that the temporal profiles of the laser pulses are chosen to be trapezoidal is only to confirm exactly our theoretical predictions from Eqs. (\ref{Eq:rate})-(\ref{Eq:He}). The regime can be applied for Gaussian or other temporal profiles of laser pulse as well by just more carefully matching the condition. The simulations of ultrathin Al foils irradiated by an intense laser pulse with $\sin^2$ temporal profile as $\sin^2(\pi t/\tau)$ have also been run, where the laser amplitude $a_0=24$ is taken and $\tau=30T_0$ is chosen to keep the total laser energy is the same as above. The final energy spectra when the laser is over are shown in Fig. \ref{fig:Alspectra}(c) and \ref{fig:Alspectra}(d) as the dashed lines, which show clearly in our regime with the ionization effect, a high-energy $\text{Al}^{13+}$ ion beam at peak energy $45\ \text{MeV}/\text{u}$ is also obtained, which has a well-peaked narrow-band energy spectrum. Even so, the high contrast of the laser in excess of $10^{10}$ is required for the regime, which can be achieved by using either plasma mirrors or parametric amplifications.

Furthermore, for heavy ion acceleration, except the ionization physics,  the bremsstrahlung radiation loss might also play roles. The bremsstrahlung radiation power formula can be estimated as \cite{Huba}
\begin{equation}
P_{Br}=1.69\times10^{-32}n_eT_e^{1/2}\sum[\zeta^2n_{\zeta}]\ \text{W}/\text{cm}^3.
\end{equation}
For ultrathin high-Z foil targets, where the foil thickness $l$ is only 10s of nm, the total radiation loses is about $E_{Br}=P_{Br}\times{r_0^2}\times{l}\times\tau\sim3.0\times10^{-6}\ \text{J}$, which is much smaller than the incident intense laser energy, 
where $E_{laser}=I\times{r_0^2}\times\tau\sim6.0\ \text{J}$. Thus, the bremsstrahlung radiation loss can be neglected in our regime.

\textit{Summary --} In summary, we have proposed a novel RPA regime for generation of high-energy mono-energetic heavy ion beams from ultrathin foils irradiated by intense laser pulses. In this regime, the ionization dynamics of high-Z elements play a key role. By matching the laser intensity with the large ionization energy gap during the successive ionization when passing the nobel gas configurations, a self-organized, stable RPA of highly-charged heavy ion beam close to the central laser axis is obtained. The ionization also provides a large number of supplementary electrons accompany and stabilize the RPA of heavy ion beams. The regime has been confirmed by our comprehensive 2D PIC simulations, which shows a highly-charged monoenergetic $\text{Al}^{13+}$ ion beam with peak energy at $1\ \text{GeV}$ and energy spread of only $5\%$ is produced by ultraintense laser at intensity $7\times10^{20}\ \text{W/}\text{cm}^{2}$ through the regime. Similarly, a quasi-monoenergetic $\text{Fe}^{26+}$ ion beam with peak energy $300\ \text{MeV}/\text{u}$ ($17\ \text{GeV}$) can be also obtained by increasing the intensity up to $10^{22}\ \text{W}/\text{cm}^2$. Both of them are experimental available with existing laser systems.


{}

\end{document}